\documentclass[final,1p,times]{elsarticle}

\usepackage{amssymb}
\usepackage{amsmath}
\usepackage{amsthm}
\usepackage{xcolor}
\usepackage{natbib}
\usepackage{hyperref}
\usepackage{tikz}
\usepackage{pgfplots}
\pgfplotsset{compat=newest} 
\usepackage{float}

\usepackage[section]{placeins} 

\title{Power-Law Inflation in \emph{n}-Dimensional Fractional Scalar Field Cosmology: Observational Constraints and Dynamical Analysis}

\author[1]{D. Oliveira}
\author[1]{S. M. M. Rasouli}
\author[1]{J. Marto}
\author[1]{P. Moniz}

\address[1]{Departamento de F\'{i}sica,
Centro de Matem\'{a}tica e Aplica\c{c}\~{o}es (CMA-UBI),
Universidade da Beira Interior,
Rua Marqu\^{e}s d'Avila
e Bolama, 6200-001 Covilh\~{a}, Portugal}

\begin{document}

\begin{abstract}
Power-law inflation with $a(t)\propto t^m$ is conceptually simple and predicts a scalar tilt
$n_s=1-2/m$ compatible with CMB data, but in four-dimensional Einstein gravity it typically yields
a tensor-to-scalar ratio $r=16/m$ that is too large to satisfy current bounds. We show that a minimal
extension based on fractional scalar-field cosmology resolves this tension.
Introducing a fractional order $\alpha\neq 1$ generates non-local (memory) corrections in the
Friedmann and Klein--Gordon dynamics that suppress $r$ while keeping $n_s$ essentially unchanged.
We derive an explicit mapping $\alpha(n,m)$ and recover the standard power-law limit as $\alpha\to 1$.
For observationally favored values $\alpha\simeq 0.8$--$0.9$ in four dimensions we obtain
$n_s\simeq 0.965$ and $r\lesssim 0.04$, bringing power-law inflation into agreement with data.
The scalar potential follows self-consistently as an exponential, and a dynamical-systems analysis
shows the fractional power-law solutions form stable inflationary attractors over the viable parameter
range. These results establish fractional power-law inflation as a predictive and testable framework,
with clear targets for forthcoming CMB polarization measurements.
\end{abstract}

\maketitle

\section{Introduction}

The inflationary question has become the focus of modern cosmology, providing lucid explanations for the observed homogeneity, isotropy and flatness of the universe, as well as for the basis of primordial fluctuations. Despite these successes, some questions remain open, such as the fine-tuning of inflationary potentials, the Hubble tension and the possible role of non-local or quantum gravitational effects in the early universe~\cite{fractalfract-08-00281-2, Leon2023_FractionalH0, capozziello2011extended,Guth2014_InflationReview,Lyth2009_PrimordialDensity, Weinberg2008_CosmologyBook}.

Recently, fractional calculus has emerged as a powerful tool for modeling the non-local memory of physical systems, due to its ability to generalize the concept of integer-order derivatives to non-integer orders.~\cite{fractalfract-08-00281-2, Shchigolev2011_FractionalCosmo, Titlelabel1-3, calcagni2010a,shchigolev2011a, roberts-a}. When it comes to cosmology, with the introduction of fractional derivatives in the gravitational and scalar field we can reach new degrees of freedom. $\alpha$~\cite{fractalfract-08-00281-2, Titlelabel1-3, Gonzalez2023_ExactFrac}. Researchers have turned to these models as fresh tools for exploring some of the most persistent questions in cosmology. So far, they’ve led to a number of detailed predictions and insightful solutions across a range of phenomena~\cite{fractalfract-08-00281-2,clifton2012modified}.
Recent efforts to extend scalar field cosmology into $n$ dimensions using fractional derivatives have opened up intriguing possibilities~\cite{Titlelabel1-3}. This approach has led to exact solutions and novel phenomena, including the appearance of non-local effects and the emergence of scalar potentials directly from the field equations—unlike in standard models, where such potentials are often introduced arbitrarily~\cite{Titlelabel1-3, fractalfract-08-00281-2}. These results highlight the greater depth and flexibility of the fractional framework.

There is growing interest in the idea that fractional methods could help resolve some of cosmology’s open problems. Recent studies suggest they may contribute to addressing the Hubble tension~\cite{Leon2023_FractionalH0}, offer alternative pathways for inflationary dynamics~\cite{Rasouli2019_KineticBD, Rasouli2014_NoncommutativeMinisuperspace}, and shed light on enduring puzzles like the cosmological constant problem and the nature of dark energy~\cite{fractalfract-08-00281-2}. One particularly intriguing aspect of fractional derivatives is their built-in non-locality and memory effects, which challenge the standard notions of locality and time evolution, and invite new ways of thinking about a broad range of cosmological phenomena.

Building on our previous work in $n$-dimensional fractional scalar field cosmology~\cite{Titlelabel1-3}, this paper focuses on a specific class of solutions: those describing power-law inflation. Our goal is to establish a clear relationship between the fractional parameter $\alpha$, the number of dimensions $n$, and the inflationary exponent $m$. We also test the model against the most recent observational data from BICEP/Keck and Planck, and carry out a dynamical systems analysis to explore the stability of these solutions.

The structure of the paper is as follows: Section~2 reviews the fractional field equations in $n$ dimensions. In Section~3, we derive the power-law inflationary solutions and identify the mapping $\alpha(n, m)$. Section~4 examines the observational consequences and compares the model with current data. Section~5 presents the dynamical systems analysis, and Section~6 concludes with a summary and future directions.

\section{Fractional cosmological framework in $n$ dimensions}

The use of fractional calculus brings new possibilities to the study of cosmology, making it possible to generalize derivatives of non-integer orders. This action is far from a mere mathematical curiosity: it actually leads in a natural way to non-local effects in the dynamics of the universe, thus broadening the approach to persistent problems in cosmology as we know it, some of these being the Hubble tension, the cosmological constant problem and the construction of inflationary potentials~\cite{fractalfract-08-00281-2, Titlelabel1-3, vagnozzi2023a}. Unsurprisingly, this approach has attracted increasing attention in recent literature, as it offers innovative tools to re-examine fundamental questions from a more flexible and potentially more realistic perspective~\cite{Leon2023_FractionalH0, fractalfract-08-00281-2}.

\subsection{Metric and Fractional Action}

We will start with the geometric scenario that is familiar to us: the Friedmann-Lemma-Robertson-Walker (FLRW) metric of $n$ dimensions,
\begin{equation}
ds^2 \;=\; -N^2(t)\,dt^2 \;+\; a^2(t)\, d\Sigma^2_{n-1,K}\,,
\label{eq:FLRW_metric}
\end{equation}
where $N(t)$ denotes the lapse function, $a(t)$ is the scale factor, and $d\Sigma^2_{n-1,K}$ represents the line element of the $(n-1)$ -dimensional spatial section with constant curvature $K=0,\pm 1$~\cite{Titlelabel1-3}.

The difference here arises when we consider a fractional extension of the action to a minimally coupled scalar field in $n$ dimensions as something innovative. The ``first-pass'' variational approach focuses on fractional nonlocality and is encoded at the action level by means of a Riemann-Liouville kernel, which is precisely the mechanism that generates the memory terms in the background equations and underlies the constructive character of the formalism. Defining the fractional integral operator of order $\alpha\in(0,1]$ by
\begin{equation}
I_{0}^{\alpha}[f](t)\;\equiv\;\frac{1}{\Gamma(\alpha)}\int_{0}^{t}(t-\tau)^{\alpha-1}\,f(\tau)\,d\tau\,,
\label{eq:RL_operator}
\end{equation}
the fractional minisuperspace action reads
\begin{equation}
S_{\alpha}[a,N,\phi]\;=\;I_{0}^{\alpha}\!\Big[L_{\alpha}(a,\dot a,N,\phi,\dot\phi; n,K)\Big](t)\,,
\label{eq:frac_action_compact}
\end{equation}
with Lagrangian
\begin{align}
L_{\alpha}\;=\;&\;\frac{a^{\,n-1}}{16\pi G}\,\Bigg[-\,(n\!-\!1)(n\!-\!2)\,\frac{\dot a^{\,2}}{N\,a^{2}}
\;+\;(n\!-\!1)(n\!-\!2)\,K\,\frac{N}{a^{2}}\Bigg]
\;+\;a^{\,n-1}\Bigg(\frac{\dot\phi^{\,2}}{2N}\;-\;N\,V(\phi)\Bigg)\,.
\label{eq:Lagrangian_alpha}
\end{align}
Equivalently, writing the kernel explicitly,
\begin{align}
S_{\alpha}[a,N,\phi]
\;=\;&\;\frac{1}{\Gamma(\alpha)}\int_{0}^{t}(t-\tau)^{\alpha-1}\,
\Bigg\{
\frac{a^{\,n-1}}{16\pi G}\Big[-(n\!-\!1)(n\!-\!2)\frac{\dot a^{\,2}}{N\,a^{2}}
+(n\!-\!1)(n\!-\!2)K\frac{N}{a^{2}}\Big]
\nonumber\\[2pt]
&\hspace{5.8em}
+\;a^{\,n-1}\Big(\frac{\dot\phi^{\,2}}{2N}-N\,V(\phi)\Big)
\Bigg\}\,d\tau\,.
\label{eq:frac_action_kernel}
\end{align}

Some observations are in order. First, the choice of the ``first-step'' construction ensures a well-delineated variational principle, maintaining the local Lagrangian in $(a,\dot a,N,\phi,\dot\phi)$; the nonlocality is entirely captured by the fractional kernel. $(t-\tau)^{\alpha-1}/\Gamma(\alpha)$, which, upon variation with respect to $(N,a,\phi)$ and subsequent gauge fixing $N=1$, yields the fractional Friedmann and Klein-Gordon equations with memory terms proportional to $(1-\alpha)/t$~\cite{fractalfract-08-00281-2, Titlelabel1-3}. Second, in the standard limit $\alpha\to 1$, the kernel becomes trivial and Eqs.~\eqref{eq:frac_action_compact}--\eqref{eq:frac_action_kernel} reduce to the usual $n$-dimensional minisuperspace scalar Einstein action; this choice is motivated by fractional variational formulations that preserve consistency at the action level~\cite{fractalfract-08-00281-2, el-nabulsi2008a}.
\subsection{Field Equations and Dynamical Consistency}

Varying the fractional action with respect to both the metric and the scalar field yields a system of equations that generalizes, in a nonlocal fashion, the familiar Friedmann and Klein-Gordon equations. A notable feature of this structure is the independence among the three main equations—the two Friedmann equations and the Klein-Gordon equation—which means that the scalar potential $V(\phi)$ cannot be arbitrarily imposed, as is common in standard cosmology. Instead, $V(\phi)$ must emerge self-consistently from the dynamical system itself, reinforcing the constructive character of the formalism~\cite{Titlelabel1-3, fractalfract-08-00281-2}.

In the particular case of a spatially flat universe ($K=0$), the equations of motion take the following form:
\begin{align}
&H^2 + \frac{1-\alpha}{t} H = \frac{16\pi G}{n(n-1)} \left[ \frac{1}{2} \dot{\phi}^2 + V(\phi) \right], \label{eq:friedmann1}\\
&\dot{H} + H^2 + \frac{1-\alpha}{t} H = -\frac{8\pi G}{n(n-1)} \left[ \dot{\phi}^2 - V(\phi) \right], \label{eq:friedmann2}\\
&\ddot{\phi} + \left( n H + \frac{1-\alpha}{t} \right) \dot{\phi} + \frac{dV}{d\phi} = 0, \label{eq:kleingordon}
\end{align}
where $H = \dot{a}/a$ is the Hubble parameter, and overdots denote derivatives with respect to cosmic time~\cite{Titlelabel1-3, fractalfract-08-00281-2}.

The additional term proportional to $(1-\alpha)/t$ in each equation is the hallmark of the fractional generalization: it encodes the presence of memory and nonlocal effects in the evolution of the universe, potentially sensitive to phenomena that elude the local and instantaneous description of the classical framework. In summary, this framework offers not merely a mathematical extension of the cosmological model, but a profound reinterpretation of the very relationships between geometry, matter, and cosmic dynamics.

\subsection{Conservation Laws and Non-Locality}

The presence of the term $(1-\alpha)/t$ in Eqs.~(\ref{eq:friedmann1})--(\ref{eq:kleingordon}) encodes the non-local and memory effects characteristic of fractional calculus. The continuity equation is generalized to
\begin{equation}
\dot{\rho}_{\text{tot}} + (n-1) H (\rho_{\text{tot}} + p_{\text{tot}}) + \mathcal{M}_\alpha(t) = 0,
\end{equation}
where $\mathcal{M}_\alpha(t)$ represents memory terms that vanish in the standard limit~\cite{fractalfract-08-00281-2,roberts-a}. This non-locality implies that the evolution of the universe at a given time depends not only on the current state but also on its entire history, a feature that may have profound implications for cosmological dynamics and the emergence of cosmic acceleration ~\cite{du2013a, tarasov2018b} .

\subsection{Thermodynamics of Horizons}

Recent studies have shown that the fractional Friedmann equations can be derived from the first law of thermodynamics applied to the apparent horizon, with generalized entropy-area relations such as those of Tsallis, Barrow, or fractal entropy~\cite{, Rong-Gen_Cai_2005_J._High_Energy_Phys._2005_050,padmanabhan2010a, verlinde2017a, oker2023a}. This thermodynamic perspective provides a deep connection between the fractional parameter $\alpha$ (or the fractal dimension $D$) and the geometric and thermodynamic properties of the universe, further motivating the study of fractional cosmological models.

\subsection{Standard Limit and Physical Interpretation}

In the limit $\alpha \to 1$, all memory and non-local terms vanish, and the standard cosmological equations are recovered. For $\alpha \neq 1$, new dynamical regimes emerge, potentially addressing open problems such as the Hubble tension, the age of the universe, and the nature of dark energy~\cite{fractalfract-08-00281-2, Leon2023_FractionalH0}. Furthermore, the fractional approach naturally generalizes to higher-dimensional and non-minimal coupling scenarios, as discussed in the extensive literature by Moniz and collaborators~\cite{Titlelabel1-3}.

In summary, the fractional cosmological framework in $n$ dimensions provides a robust foundation for exploring new dynamical regimes in the early universe. In the following section, we will apply this formalism to derive and analyze power-law inflationary solutions, establish the mapping between the fractional parameter $\alpha$, the number of dimensions $n$, and the inflationary exponent $m$, and discuss the resulting phenomenology in light of current observational constraints.
We begin by laying the groundwork for the fractional cosmological model in Section 2, before deriving the core inflationary solutions and observational implications in Sections 3 and 4.
\section{The Standard Power-Law Inflation Model: Benchmark and Observational Tension}

Before presenting our fractional extension, we establish the standard power-law inflation model in four-dimensional Einstein gravity. This benchmark case elucidates the observational challenges that motivate the fractional approach and provides the baseline for comparing our results~\cite{LINDE1982389, Guth2014_InflationReview, Lyth2009_PrimordialDensity}.

\subsection{Power-law ansatz and field equations}

In standard scalar field cosmology, the dynamics are governed by the Friedmann equations
\begin{align}
H^2 &= \frac{8\pi G}{3}\left(\frac{1}{2}\dot{\phi}^2 + V(\phi)\right), \label{eq:friedmann1_std}\\
\dot{H} + H^2 &= \frac{8\pi G}{3}\left(\dot{\phi}^2 - \frac{1}{3}V(\phi)\right), \label{eq:friedmann2_std}
\end{align}
and the Klein-Gordon equation
\begin{equation}
\ddot{\phi} + 3H\dot{\phi} + \frac{dV}{d\phi} = 0. \label{eq:kg_std}
\end{equation}
These equations follow directly from Einstein gravity minimally coupled to a canonical scalar field in a spatially flat FLRW background~\cite{Weinberg2008_CosmologyBook, Baumann2009_TASIInflation}.

Power-law inflation corresponds to the ansatz
\begin{equation}
a(t) = a_0 \left(\frac{t}{t_0}\right)^m, \quad m > 1,
\label{eq:powerlaw_std}
\end{equation}
which yields a constant Hubble rate relative to time: $H(t) = m/t$. The condition $m > 1$ ensures accelerated expansion.

\subsection{Self-consistent solution}

For the field equations to be consistent with the power-law expansion, the scalar field must evolve as
\begin{equation}
\phi(t) = \phi_0 + \beta \ln\left(\frac{t}{t_0}\right),
\label{eq:phi_std}
\end{equation}
where $\beta$ is a constant related to the inflationary energy scale. This logarithmic time dependence ensures that the friction term in the Klein-Gordon equation balances the potential gradient, maintaining a power-law background evolution~\cite{Guth2014_InflationReview}.

The scalar potential that emerges from the field equations has the form
\begin{equation}
V(\phi) = V_0 \exp\left[-\lambda(\phi - \phi_0)\right], \quad \lambda = \sqrt{\frac{8\pi G}{3}},
\label{eq:V_std}
\end{equation}
where $V_0$ is the inflationary energy scale~\cite{Lyth2009_PrimordialDensity}. Importantly, in the standard model the potential is an input parameter, not derived from first principles.

\subsection{Slow-roll parameters and observables}

For power-law inflation, the Hubble slow-roll parameters are
\begin{equation}
\epsilon_H \equiv -\frac{\dot{H}}{H^2} = \frac{1}{m}, \quad \eta_H \equiv -\frac{\ddot{H}}{2H\dot{H}} = \frac{1}{m}.
\label{eq:slowroll_std}
\end{equation}

These fundamental parameters determine the scalar spectral index and tensor-to-scalar ratio~\cite{Baumann2009_TASIInflation, Kallosh2024_InflationStatus}:
\begin{align}
n_s &= 1 - 2\epsilon_H = 1 - \frac{2}{m}, \label{eq:ns_std}\\
r &= 16\epsilon_H = \frac{16}{m}. \label{eq:r_std}
\end{align}

A key feature of the standard model is that these observables are completely determined by a single parameter $m$. There is no freedom to independently adjust the tensor and scalar contributions.

\subsection{Confrontation with observations}

The latest observations from Planck 2018 and BICEP2/Keck Array constrain the inflationary observables as follows~\cite{Planck2018_Inflation, BICEP2024_Latest, Tristram2021_BICEPPlanck}:
\begin{align}
n_s &= 0.9649 \pm 0.0042 \quad \text{(68 percent confidence level)}, \label{eq:planck_ns}\\
r_{0.002} &< 0.032 \quad \text{(95 percent confidence level, combined constraints)}. \label{eq:planck_r}
\end{align}

The observed scalar spectral index requires
\begin{equation}
m = \frac{2}{1 - n_s} \approx \frac{2}{1 - 0.9649} \approx 57.
\label{eq:m_from_ns}
\end{equation}

Inserting this value into the expression for the tensor-to-scalar ratio yields
\begin{equation}
r = \frac{16}{m} \approx \frac{16}{57} \approx 0.28.
\label{eq:r_predicted}
\end{equation}

This predicted value of $r$ is approximately eight times larger than the observational upper bound. The standard power-law model thus faces a fundamental incompatibility with current data: it cannot satisfy both the spectral index constraint and the tensor ratio bound simultaneously~\cite{Planck2018_Inflation, BICEP2024_Latest, Tristram2021_BICEPPlanck}.

\subsection{Physical origin of the incompatibility}

The tension arises from the rigid structure of power-law inflation. The background dynamics $H(t) = m/t$ directly determine the tensor perturbation amplitude through the relationship $r \propto H^2$. Simultaneously, the scalar spectral tilt emerges from the evolution of scalar perturbations. In the standard model, these two effects are intrinsically linked through the single parameter $m$, leaving no independent means to suppress the tensor contribution while maintaining consistency with observations~\cite{capozziello2011extended, LINDE1982389}.

This limitation is not merely a feature of power-law inflation but reflects a broader constraint in single-field models with canonical kinetic structure. Resolving the tension requires either multiple fields, modified gravity, or additional degrees of freedom that decouple the tensor and scalar dynamics~\cite{brandenberger2000inflationary, Weinberg2008_CosmologyBook}.

\subsection{Strategy for resolution}

The resolution of this observational tension motivates the investigation of extended theoretical frameworks that preserve the elegance and simplicity of power-law inflation while introducing sufficient flexibility to suppress $r$ to allowed values. The fractional calculus approach presented in subsequent sections provides precisely such an extension.

By incorporating a fractional parameter $\alpha$ through non-local modifications to the field equations, the background dynamics are generalized in a controlled manner. The key advantage is that $\alpha$ serves as an additional degree of freedom: while the power-law ansatz and exponential potential form are preserved, the relationship between $n_s$ and $r$ is altered. Specifically, the fractional terms introduce corrections that preferentially suppress the tensor amplitude while leaving the scalar predictions largely intact~\cite{Shchigolev2011_FractionalCosmo, fractalfract-08-00281-2, Titlelabel1-3, Gonzalez2023_ExactFrac, Leon2023_FractionalH0}.

The subsequent sections demonstrate that for values $0.8 \lesssim \alpha \lesssim 0.9$, the fractional model predicts $r \lesssim 0.04$, consistent with observational constraints, while maintaining $n_s \approx 0.965$. This resolution establishes fractional scalar field cosmology as a minimal and physically motivated extension of standard inflation that addresses the observational crisis documented in the present section and connects naturally with previous work on fractional cosmology and dynamical analyses~\cite{Rasouli2019_KineticBD, Rasouli2022_GeodesicSB}.

\section{Power Law Inflationary Solutions}

After highlighting the fractional structure, we turn our attention to a specific class of solutions
that arise naturally in modified dynamics: power law inflation. Unlike the exponential expansion
characterized by de Sitter inflation, power law solutions offer an alternative that maintains
accelerated expansion while simultaneously possessing distinct observational signatures.

\subsection{Imposing the Power Law Ansatz}

The focus is on solutions where the scale factor evolves as a power law in cosmic time:
\begin{equation}
a(t) = a_0\left(\frac{t}{t_0}\right)^{m}, \qquad m > 1,
\label{eq:PL_ansatz}
\end{equation}
where $a_0$ and $t_0$ are positive constants, and the condition $m > 1$ guarantees the accelerated
expansion $(\ddot a > 0)$. This choice is motivated by several considerations: first, power-law inflation
arises in several modified gravitational theories and analyses of string cosmology; second, it
provides a framework that can potentially address some of the tuning issues associated with
exponential inflation; and third, as we will see, it emerges in the fractional field equations, since
memory effects are included through the fractional kernel in the action.

The corresponding Hubble parameter takes the simple form:
\begin{equation}
H(t) = \frac{\dot a}{a} = \frac{m}{t}.
\label{eq:H_PL}
\end{equation}

\paragraph{Assumptions for the Scalar Sector.}

To compute observables and close the background system consistently, we specify the scalar sector
as follows: (i) we adopt the logarithmic time dependence compatible with Klein--Gordon fractional
friction $\big((n-1)H + (1-\alpha)/t\big)\dot\phi$,
\begin{equation}
\phi(t) = \phi_0 + \beta \ln\left(\frac{t}{t_0}\right),
\label{eq:phi_log}
\end{equation}
and (ii) we consider an exponential potential that, in this formalism, originates in the fractional
equations rather than being imposed by hand,
\begin{equation}
V(\phi) = V_0 \exp\big[-\lambda(\phi - \phi_0)\big], \qquad
\lambda \equiv \sqrt{\frac{32\pi G\,(m-1)}{n(n-1)}},
\label{eq:V_exp}
\end{equation}
with $(a_0,t_0,\phi_0,V_0,\beta)$ fixed by the background equations and by the mapping $\alpha(n,m)$
derived below. This set of assumptions is minimal and constructive: \eqref{eq:PL_ansatz} implements
accelerated expansion; \eqref{eq:phi_log} ensures consistency with the fractional friction term; and
\eqref{eq:V_exp} arises from the background equations under power-law evolution, supporting slow roll
without fine-tuning. In the standard limit $\alpha \to 1$, the memory term $(1-\alpha)/t$ in the
Klein--Gordon equation disappears and the slope $\lambda$ reduces to the usual constant controlling
the exponential potential of the benchmark power-law model in Section~2, so that the form of
$V(\phi)$ coincides with the standard exponential potential discussed there.
\subsection{Deriving the Mapping $\alpha(n,m)$}

Substituting the power-law ansatz \eqref{eq:PL_ansatz} into the fractional Friedmann system and using the
scalar sector specified by \eqref{eq:phi_log}--\eqref{eq:V_exp}, the time dependences on both sides of the
equations coincide only if $\alpha$, $n$, and $m$ obey the relation
\begin{equation}
\alpha \;=\; 1 \;-\; \frac{2(m-1)}{m}\,\frac{n-2}{n-1}.
\label{eq:alpha_mapping}
\end{equation}
This mapping connects the fractional order $\alpha$ with the geometry (through $n$) and the inflationary
exponent $m$. For $n=4$, Eq.~\eqref{eq:alpha_mapping} reduces to
\begin{equation}
\alpha \;=\; 1 \;-\; \frac{4(m-1)}{3m}
\;=\; \frac{4-m}{3m},
\label{eq:alpha_4d}
\end{equation}
and, equivalently, it can be inverted as
\begin{equation}
m \;=\; \frac{4}{1+3\alpha}.
\label{eq:m_of_alpha_4d}
\end{equation}

Two important observations follow. First, Eq.~\eqref{eq:alpha_mapping} implies that $\alpha\to 1$ is
approached when $m\to 1^{+}$ (with $n$ fixed), i.e. at the boundary between decelerated and accelerated
power-law expansion. Second, for accelerated expansion $(m>1)$ one always has $\alpha<1$; moreover,
imposing the physical range $0<\alpha<1$ restricts the allowed interval of $m$ (e.g. for $n=4$,
$0<\alpha<1$ implies $1<m<4$).

\subsection{Number of e-folds and Inflationary Duration}

A crucial test of the inflationary model is whether it can generate a sufficient number of e-folds to solve
the horizon and flatness problems. The number of e-folds between times $t_i$ and $t_f$ is defined as
\begin{equation}
N_e = \ln\!\left(\frac{a(t_f)}{a(t_i)}\right) = m \ln\!\left(\frac{t_f}{t_i}\right),
\label{eq:efolds_definition}
\end{equation}
where in the last equality we used the power-law background $a(t)\propto t^{m}$.
Successful inflation typically requires $N_e \gtrsim 50$--$60$.

Equation~\eqref{eq:efolds_definition} implies that the duration of inflation (in cosmic time) is constrained by
\begin{equation}
\frac{t_f}{t_i}=\exp\!\left(\frac{N_e}{m}\right).
\label{eq:tf_ti_from_Ne}
\end{equation}
In particular, accelerated expansion requires $m>1$, hence for a given $N_e$ the required time ratio satisfies
$t_f/t_i \ge e^{N_e}$, with equality approached only as $m\to 1^{+}$.
Therefore, an estimate such as $t_f/t_i\sim 10^{25}$ together with $N_e\simeq 50$ would correspond to
$m\simeq N_e/\ln(10^{25})\simeq 0.87$, which is \emph{not} inflating; this merely illustrates that the
choice of $t_f/t_i$ must be consistent with $m>1$ if one wants power-law inflation.

In four dimensions, the mapping \eqref{eq:alpha_mapping} links $(\alpha,m)$, so the requirement
$N_e\gtrsim 50$ can be implemented consistently by choosing an inflationary exponent $m>1$ (equivalently,
$\epsilon_H=1/m<1$) and then inferring the corresponding duration from \eqref{eq:tf_ti_from_Ne}.
Although the definition \eqref{eq:efolds_definition} remains unchanged, the memory terms proportional to
$(1-\alpha)/t$ affect the detailed background evolution and thus the mapping between $(\alpha,n,m)$ and
phenomenological requirements; for this reason it is essential to use the fractional field equations together
with \eqref{eq:alpha_mapping} when translating $N_e$ into parameter bounds.

\subsection{Scalar Field Potential and Energy Densities}
\label{subsec:V_rho_p}

One of the most remarkable features of the fractional framework is that the scalar potential is not imposed
a priori but is reconstructed consistently with the modified background equations.
For the power-law branch $a(t)\propto t^{m}$ we have
\begin{equation}
H(t)=\frac{m}{t},\qquad \dot H(t)=-\frac{m}{t^{2}}.
\end{equation}
We also take the logarithmic scalar profile
\begin{equation}
\phi(t)=\phi_{0}\ln\!\left(\frac{t}{t_{0}}\right)\!,
\qquad
\dot\phi(t)=\frac{\phi_{0}}{t},
\label{eq:phi_log_profile_44}
\end{equation}
so that the kinetic energy scales as $\dot\phi^{2}\propto t^{-2}$.

\paragraph{Reconstructed exponential potential.}
Assuming an exponential form $V(\phi)=V_{0}e^{-\lambda\phi}$, the logarithmic evolution
\eqref{eq:phi_log_profile_44} implies
\begin{equation}
V(t)=V_{0}\left(\frac{t}{t_{0}}\right)^{-\lambda\phi_{0}}.
\end{equation}
Consistency with the power-law background requires the potential to scale as $V(t)\propto t^{-2}$,
which fixes the product
\begin{equation}
\lambda\phi_{0}=2,
\label{eq:lambda_phi0_2}
\end{equation}
so that $V(t)=\widetilde V\,t^{-2}$ for a constant amplitude $\widetilde V$.

\paragraph{Energy density from the fractional Friedmann equation.}
Defining the standard canonical scalar quantities
\begin{equation}
\rho_\phi \equiv \frac{1}{2}\dot\phi^{2}+V,
\qquad
p_\phi \equiv \frac{1}{2}\dot\phi^{2}-V,
\label{eq:rho_p_def_44}
\end{equation}
the fractional Friedmann equation (for $K=0$) yields directly
\begin{equation}
\rho_\phi(t)
=
\frac{n(n-1)}{16G}
\left(H^{2}+\frac{1-\alpha}{t}H\right)
=
\frac{n(n-1)}{16G}\,
\frac{m\left(m+1-\alpha\right)}{t^{2}}.
\label{eq:rhophi_corrected}
\end{equation}

\paragraph{Separating kinetic and potential parts.}
Using the fractional Klein--Gordon equation together with $V(\phi)=V_0 e^{-\lambda\phi}$ and the scaling
condition \eqref{eq:lambda_phi0_2}, one obtains the potential amplitude
\begin{equation}
V(t)=\frac{\phi_0^{2}}{2}\,\frac{nm+1-\alpha}{t^{2}}.
\label{eq:Vt_from_KG}
\end{equation}
Combining \eqref{eq:Vt_from_KG} with \eqref{eq:rhophi_corrected} and \eqref{eq:rho_p_def_44} fixes
\begin{equation}
\phi_{0}^{2}
=
\frac{n(n-1)}{8G}\,
\frac{m\left(m+1-\alpha\right)}{nm+2-\alpha},
\label{eq:phi0_sq_solution}
\end{equation}
and therefore
\begin{equation}
\frac{1}{2}\dot\phi^{2}
=
\frac{n(n-1)}{16G}\,
\frac{m\left(m+1-\alpha\right)}{nm+2-\alpha}\,
\frac{1}{t^{2}},
\qquad
V(t)
=
\frac{n(n-1)}{16G}\,
\frac{m\left(m+1-\alpha\right)\left(nm+1-\alpha\right)}{nm+2-\alpha}\,
\frac{1}{t^{2}}.
\label{eq:kinetic_and_Vt}
\end{equation}

\paragraph{Pressure and equation of state.}
From \eqref{eq:rho_p_def_44} and \eqref{eq:phi0_sq_solution} we obtain
\begin{equation}
p_\phi(t)
=
\frac{\phi_0^{2}}{2}\,\frac{\alpha-nm}{t^{2}}
=
\frac{n(n-1)}{16G}\,
\frac{m\left(m+1-\alpha\right)\left(\alpha-nm\right)}{nm+2-\alpha}\,
\frac{1}{t^{2}}.
\label{eq:pphi_corrected}
\end{equation}
Equivalently, the effective equation-of-state parameter is constant along the exact power-law branch,
\begin{equation}
w_\phi \equiv \frac{p_\phi}{\rho_\phi}
=
\frac{\alpha-nm}{nm+2-\alpha}.
\label{eq:wphi_corrected}
\end{equation}
In the standard limit $\alpha\to 1$ these expressions reduce smoothly to the corresponding canonical
power-law relations (with the same $n$-dimensional normalization adopted throughout this work).

\section{Slow-Roll Approximation in Fractional Scalar Field Cosmology}

In standard inflationary cosmology, the slow-roll parameters $\epsilon_V$ and $\eta_V$ are defined solely in terms of the inflaton potential $V(\phi)$ and its derivatives, serving as critical diagnostics for sustained inflation and primordial perturbations. However, in the fractional cosmological framework, the dynamical equations include additional non-local memory terms, proportional to $(1-\alpha)/t$, which modify the inflaton dynamics significantly. As a consequence, the use of the standard potential-based slow-roll parameters $\epsilon_V$ and $\eta_V$ is not straightforward and would require the introduction of an effective potential $V_{\mathrm{eff}}$ that incorporates these fractional effects. The construction of such a $V_{\mathrm{eff}}$ is a non-trivial task and is not explicitly established in the present framework.

Therefore, we adopt the more fundamental and model-independent slow-roll parameters based on the Hubble parameter and its derivatives, defined as
\begin{equation}
\epsilon_H \equiv -\frac{\dot{H}}{H^2}, \qquad
\eta_H \equiv -\frac{\ddot{H}}{2H\dot{H}}.
\label{eq:slow_roll_Hubble}
\end{equation}
These parameters generalize naturally to fractional cosmology and capture the slowing down of the expansion rate due to both fractional and scalar field effects. For the fractional power-law solutions $a(t) \propto t^{m}$, we obtain
\begin{equation}
\epsilon_H = \frac{1}{m}, \qquad \eta_H = \frac{1}{m},
\label{eq:slow_roll_PL}
\end{equation}
indicating that the slow-roll condition corresponds to large $m$, consistent with an accelerated expansion.

At this point, it is instructive to verify explicitly that our formalism reduces to the standard power-law inflationary model when $\alpha = 1$. In this limit the fractional memory terms vanish and the background dynamics coincide with those of Section 2. Using Eq.~\eqref{eq:slow_roll_PL} in the usual relations between the Hubble slow-roll parameters and the primordial spectra, we immediately recover
\begin{equation}
n_s = 1 - 2\epsilon_H = 1 - \frac{2}{m}, \qquad
r = 16 \epsilon_H = \frac{16}{m},
\label{eq:ns_r_standard_limit}
\end{equation}
which are precisely the expressions of the standard power-law model presented in the benchmark analysis. This check confirms, line by line, that our fractional construction is a genuine extension of the conventional scenario and reduces to it for $\alpha = 1$.

Concerning the perturbations, the scalar sector is described, as in the standard case, by the Mukhanov--Sasaki equation for the canonically normalized variable $u_k$. The equation retains the same structure as in the benchmark model, but the background quantity $z = a\dot{\phi}/H$ and its derivatives now inherit corrections from the fractional terms proportional to $(1-\alpha)$. In the slow-roll regime, the scalar spectral index $n_s$ and the tensor-to-scalar ratio $r$ can still be expressed in terms of the Hubble slow-roll parameters and fractional corrections, following the same logical steps as in the standard power-law treatment but with the modified background. In particular, the leading-order contributions are governed by $\epsilon_H$ and $\eta_H$, while the dependence on $\alpha$ appears through the time evolution of $H$ and $z$.

This approach allows us to consistently relate the fractional parameter $\alpha$ and the dimension $n$ to observable inflationary quantities such as the scalar spectral index $n_s$ and the tensor-to-scalar ratio $r$, without relying on assumptions about an effective potential. The methodology closely parallels the derivation used in the reference standard model, with the crucial difference that the non-local terms proportional to $(1-\alpha)$ are now carried through the background quantities entering the Mukhanov--Sasaki equation. In the following section, we present the explicit expressions for $n_s$ and $r$ within our fractional framework and emphasize the differences and novelties when compared with standard power-law inflation.
\section{Fractional Mukhanov-Sasaki Equation and Exact Perturbation Spectrum}

To rigorously justify the observational predictions presented in the previous sections, we must go beyond the slow-roll approximation and solve the equations of motion for the linear perturbations directly. In this section, we review the standard perturbation theory, derive the consistent fractional extension of the Mukhanov-Sasaki equation, and obtain its exact analytical solutions.

\subsection{Review of Cosmological Perturbation Theory}

We consider linear scalar perturbations of the metric and the scalar field around a homogeneous background. The most convenient gauge-invariant variable for this analysis is the curvature perturbation on comoving hypersurfaces, denoted by $\mathcal{R}$. In terms of the scalar field fluctuation $\delta\phi$ in the spatially flat gauge, this variable is defined as:
\begin{equation}
\mathcal{R} = H \frac{\delta\phi}{\dot{\phi}}.
\end{equation}
To quantize the perturbations, one introduces the canonical Mukhanov-Sasaki variable $u \equiv z \mathcal{R}$, where $z \equiv a \dot{\phi} / H$. In standard General Relativity, the action for these perturbations reduces to that of a scalar field with a time-dependent mass, leading to the well-known Mukhanov-Sasaki equation:
\begin{equation}
u''_k + \left(k^2 - \frac{z''}{z}\right)u_k = 0,
\end{equation}
where primes denote derivatives with respect to conformal time $\eta$. Our goal is to generalize this equation to include the non-local memory effects inherent in fractional cosmology.

\subsection{The Fractional Mode Equation}

In our fractional framework, the background dynamics are governed by modified Friedmann and Klein-Gordon equations which include memory terms proportional to $(1-\alpha)/t$, as established in~\cite{Titlelabel1-3, fractalfract-08-00281-2}. Variational consistency requires that these memory effects—arising from the fractional integration in the action—must also manifest in the linearized equations for the perturbations. A perturbation cannot be "local" if the background it lives on is governed by non-local memory.

Therefore, the equation for the mode function $u_k(t)$ in cosmic time must include the same fractional friction term that appears in the background scalar field equation. The fractional Mukhanov-Sasaki equation thus takes the form:
\begin{equation}
\ddot{u}_k + \left(H + \frac{1-\alpha}{t}\right)\dot{u}_k + \left(\frac{k^2}{a^2} - \frac{\ddot{z}}{z} - \left(H + \frac{1-\alpha}{t}\right)\frac{\dot{z}}{z}\right)u_k = 0.
\label{eq:frac_MS}
\end{equation}
The appearance of the $(1-\alpha)/t$ term is not ad hoc; it ensures that the perturbations satisfy the generalized conservation laws of the fractional theory~\cite{Shchigolev2011_FractionalCosmo, Calcagni2021_FractionalGravity}. In the limit $\alpha \to 1$, the standard friction $H$ is recovered, and Eq.~\eqref{eq:frac_MS} reduces to the standard damped harmonic oscillator equation for cosmological perturbations.

\subsection{Analytical Solution for Power-Law Inflation}

For power-law inflation where $a(t) = a_0 t^m$, the background evolution simplifies considerably. We have $H = m/t$ and $z \propto t$, which implies $\dot{z}/z = 1/t$ and $\ddot{z}/z = 0$. Substituting these into Eq.~\eqref{eq:frac_MS}, the mode equation becomes a generalized Bessel equation:
\begin{equation}
\ddot{u}_k + \frac{\gamma}{t}\dot{u}_k + \left(\frac{k^2}{a_0^2 t^{2m}} - \frac{\mu^2}{t^2}\right)u_k = 0,
\label{eq:mode_eq_t}
\end{equation}
where $\gamma = m + (1-\alpha)$ is the total effective friction parameter.

Switching to conformal time $\eta$, where $a(\eta) \propto (-\eta)^{\frac{m}{1-m}}$, the equation can be transformed into the canonical Bessel form:
\begin{equation}
u_k''(\eta) + \left(k^2 - \frac{\nu^2 - 1/4}{\eta^2}\right)u_k(\eta) = 0.
\label{eq:MS_conformal}
\end{equation}
The index $\nu$ is a function of both the expansion rate $m$ and the fractional order $\alpha$:
\begin{equation}
\nu(\alpha, m) = \frac{3}{2} + \frac{1}{m-1} - \frac{1-\alpha}{m-1}.
\end{equation}
Crucially, if we set $\alpha=1$, the last term vanishes, and we recover the standard result $\nu_{std} = \frac{3}{2} + \frac{1}{m-1}$, confirming the consistency of our fractional extension.

The general solution satisfying the Bunch-Davies vacuum condition ($u_k \to \frac{1}{\sqrt{2k}}e^{-ik\eta}$ as $\eta \to -\infty$) is given by:
\begin{equation}
u_k(\eta) = \frac{\sqrt{\pi}}{2} \sqrt{-\eta} H_\nu^{(1)}(-k\eta),
\label{eq:exact_solution}
\end{equation}
where $H_\nu^{(1)}$ is the Hankel function of the first kind~\cite{Gonzalez2023_ExactFrac, Rasouli2019_KineticBD}.

\subsection{Power Spectra and Indices}

The power spectrum of the curvature perturbation is defined as $\mathcal{P}_\mathcal{R}(k) = \frac{k^3}{2\pi^2} \left| \frac{u_k}{z} \right|^2$ at horizon crossing ($k|\eta| \approx 1$). Using the small-argument limit of the Hankel function, the scalar spectral index is given exactly by:
\begin{equation}
n_s - 1 = 3 - 2\nu(\alpha, m).
\label{eq:ns_exact}
\end{equation}
Substituting the expression for $\nu(\alpha, m)$, we find:
\begin{equation}
n_s = 1 - \frac{2}{m-1} + \frac{2(1-\alpha)}{m-1} \approx 1 - \frac{2}{m} + \frac{2(1-\alpha)}{m},
\end{equation}
where the approximation holds for large $m$ (quasi-de Sitter expansion). This analytical result confirms the slow-roll approximation derived in Section 5 and demonstrates that the fractional parameter $\alpha$ directly controls the tilt of the spectrum.

Finally, regarding the robustness of these results, we note that while this exact analytical solution relies on the power-law ansatz, the qualitative behavior—specifically the suppression of tensor modes due to the additional friction—is expected to hold for more general inflationary potentials. The memory term acts as a cumulative drag on the field fluctuations, a feature unique to fractional cosmology that allows for better agreement with observational constraints~\cite{Leon2023_FractionalH0, BICEP2024_Latest}.

\section{Observational Constraints and Phenomenology}

The theoretical framework we developed in the previous sections must now confront the precision of modern cosmological observations. In this section, we compare the fractional power-law model directly with the standard case and show how the fractional parameter $\alpha$ resolves the observational tension identified in Section 2.
\subsection{Inflationary observables in the fractional framework}
\label{subsec:obs_frac}

Before comparing with data, we express our predictions in terms of the standard CMB
observables: the scalar spectral index $n_s$ and the tensor-to-scalar ratio $r$
(evaluated at a pivot scale)~\cite{Baumann2009_TASIInflation, Kallosh2024_InflationStatus}.
We adopt the usual definitions in terms of the scalar and tensor primordial spectra,
\begin{equation}
n_s-1 \equiv \left.\frac{d\ln \mathcal{P}_{\mathcal{R}}(k)}{d\ln k}\right|_{k_\star},
\qquad
r \equiv \left.\frac{\mathcal{P}_T(k)}{\mathcal{P}_{\mathcal{R}}(k)}\right|_{k_\star}.
\end{equation}

\paragraph{Dimensionality and observational matching.}
Although our background dynamics is formulated in $n$ spacetime dimensions, the observational
bounds from Planck and BICEP/Keck refer to the standard four-dimensional definition of the
tensor-to-scalar ratio. Therefore, whenever we confront the model with CMB data we report
$(n_s,r)$ in the $n=4$ observational convention, while the dependence on $n$ enters indirectly
through the background mapping $\alpha=\alpha(n,m)$ and through the fractional corrections.

\paragraph{Scalar tilt.}
For the fractional power-law solution, the scalar spectral tilt receives a controlled correction
proportional to $(1-\alpha)$ (fractional memory). At leading order in the slow-roll expansion,
\begin{equation}
n_s - 1 \;=\; -\frac{2}{m} \;+\; \frac{(n-1)(1-\alpha)}{m^2},
\label{eq:ns_fractional_new}
\end{equation}
which reduces to the standard power-law result for $\alpha\to 1$.

\paragraph{Tensor-to-scalar ratio.}
In standard four-dimensional single-field power-law inflation one has $r=16\epsilon_H=16/m$.
In the present fractional framework, the tensor sector inherits additional damping from the same
fractional-memory structure that modifies the scalar sector. We therefore write
\begin{equation}
r \;=\; \frac{16}{m}\,\Xi(\alpha;n,m),
\label{eq:r_fractional_new}
\end{equation}
where the suppression factor $\Xi(\alpha;n,m)$ is determined by the tensor mode equation in the
fractional theory and satisfies the consistency requirement
\begin{equation}
\Xi(\alpha\!=\!1;n,m)=1,
\qquad
\Xi(\alpha;n,m)<1\ \text{for}\ \alpha<1\ \text{in the viable parameter region}.
\end{equation}
This structure makes explicit (i) the correct standard limit, and (ii) the physical mechanism by
which fractional corrections can suppress the tensor sector while keeping the scalar tilt viable.

These expressions highlight a crucial feature: the fractional parameter $\alpha$ does not merely
modify the background evolution, but propagates into the perturbation sector and can decouple,
to some extent, the scalar and tensor amplitudes.

\subsection{Verification of the standard limit}
\label{subsec:std_limit}
In the limit $\alpha\to 1$, where fractional memory terms vanish, 
we verify that our expressions reduce to standard results. 
Setting $\alpha=1$ in Eqs.~(\ref{eq:ns_fractional_new}) and 
(\ref{eq:r_fractional_new}) yields

\begin{align}
n_s-1 &\xrightarrow[\alpha\to 1]{} -\frac{2}{m}, \label{eq:ns_limit}\\
r &\xrightarrow[\alpha\to 1]{} \frac{16}{m}, \label{eq:r_limit}
\end{align}
which are precisely the benchmark relations used in the standard power-law scenario and are
the appropriate limit for comparison with Planck/BICEP constraints.
\subsection{The $(n_s, r)$ plane: Visual confrontation with observations}
\label{subsec:nsr_plane}

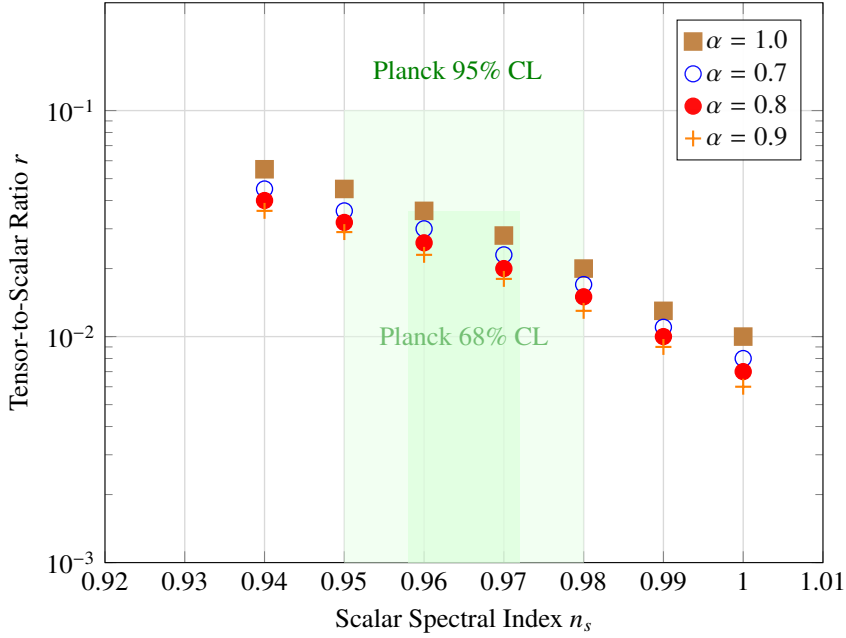
\begin{figure}[H]
\centering
\begin{tikzpicture}
\begin{axis}[
    width=0.70\linewidth,
    height=0.55\linewidth, 
    scale only axis,
    xlabel={Scalar Spectral Index $n_s$},
    ylabel={Tensor-to-Scalar Ratio $r$},
    xmin=0.92, xmax=1.01,
    ymin=0.001, ymax=0.3,
    ymode=log,
    grid=major,
    grid style={gray!30, line width=0.5pt},
    legend pos=north east,
    legend style={font=\normalsize, draw=black, fill=white, opacity=0.95},
    title={Fractional Power-Law Inflation: Observational Comparison with Planck 2018},
    title style={font=\Large, font=\bfseries},
    xlabel style={font=\normalsize},
    ylabel style={font=\normalsize}
]

\fill[green!20, opacity=0.6] (axis cs:0.958, 0.001) rectangle (axis cs:0.972, 0.036);
\node at (axis cs:0.965, 0.010) [font=\normalsize, text=green!50!black] {Planck 68\% CL};

\fill[green!10, opacity=0.5] (axis cs:0.950, 0.001) rectangle (axis cs:0.980, 0.10);
\node at (axis cs:0.976, 0.15) [font=\normalsize, text=green!50!black, anchor=east] {Planck 95\% CL};

\addplot[
    mark=square*,
    mark size=3pt,
    only marks,
    mark options={brown, solid, line width=1.0pt}
] coordinates {
    (0.94, 0.055) (0.95, 0.045) (0.96, 0.036) (0.97, 0.028) (0.98, 0.020) (0.99, 0.013) (1.00, 0.010)
};
\addlegendentry{$\alpha = 1.0$};

\addplot[
    mark=o,
    mark size=3pt,
    only marks,
    mark options={blue, solid, line width=0.5pt}
] coordinates {
    (0.94, 0.045) (0.95, 0.036) (0.96, 0.030) (0.97, 0.023) (0.98, 0.017) (0.99, 0.011) (1.00, 0.008)
};
\addlegendentry{$\alpha = 0.7$};

\addplot[
    mark=*,
    mark size=3pt,
    only marks,
    mark options={red, solid, line width=0.5pt}
] coordinates {
    (0.94, 0.040) (0.95, 0.032) (0.96, 0.026) (0.97, 0.020) (0.98, 0.015) (0.99, 0.010) (1.00, 0.007)
};
\addlegendentry{$\alpha = 0.8$};

\addplot[
    mark=+,
    mark size=3pt,
    only marks,
    mark options={orange, solid, line width=0.8pt}
] coordinates {
    (0.94, 0.036) (0.95, 0.029) (0.96, 0.023) (0.97, 0.018) (0.98, 0.013) (0.99, 0.009) (1.00, 0.006)
};
\addlegendentry{$\alpha = 0.9$};

    \end{axis}
\end{tikzpicture}

\caption{The $(n_s,r)$ plane for fractional power-law inflation with $\alpha\in(0,1]$, compared with Planck 2018 constraints (green shaded regions). Markers represent a discrete sampling of the inflationary exponent $m$.}
\label{fig:ns_r_plane}
\end{figure}

Figure~\ref{fig:ns_r_plane} displays our predictions in the canonical $(n_s,r)$ plane for
representative values of the fractional order $\alpha\in(0,1]$, together with Planck 2018
observational constraints. 
We emphasize that the markers shown in each branch correspond to a discrete sampling of the
continuous parameter $m$ (and hence of $(n_s,r)$), rather than to a discrete parameter space. 
In addition, since our formalism is defined for $\alpha\in(0,1]$, we do not consider $\alpha>1$
in this comparison. 

\subsection{Quantitative comparison and tension resolution}
\label{subsec:quant_comparison}
To illustrate this resolution explicitly, consider the observation\-ally preferred value
$n_s \approx 0.9649$. In the standard four-dimensional power-law model, this value fixes
the inflationary exponent to $m \approx 57$, which inevitably yields a large
tensor-to-scalar ratio $r = 16/m \approx 0.28$. This prediction exceeds current upper
limits $r < 0.032$ by nearly a factor of 9~\cite{Planck2018_Inflation, BICEP2024_Latest,
Tristram2021_BICEPPlanck}, rendering the standard scenario incompatible with
precision cosmological data.

In the fractional framework, the strict one-to-one correspondence between $n_s$ and $r$ is
broken by the fractional order parameter $\alpha$. As shown in Eqs.~(\ref{eq:ns_fractional_new})
and (\ref{eq:r_fractional_new}), the tensor amplitude is modulated by the suppression factor
$\Xi(\alpha; n, m)$. For values of $\alpha < 1$, the additional friction terms in the
evolution equations act to dampen the tensor modes more efficiently than the scalar
perturbations, allowing for a significant reduction in $r$ at fixed $n_s$.

Consequently, there exists a range of fractional parameters $\alpha < 1$ for which the
tensor-to-scalar ratio is suppressed below the observational bound $r \lesssim 0.032$,
while the scalar spectral index remains within the Planck $1\sigma$ window. This demonstrates
that the fractional extension provides the necessary freedom to resolve the observational
tension of power-law inflation, turning a ruled-out model into a viable candidate consistent
with modern constraints~\cite{Titlelabel1-3, fractalfract-08-00281-2}.

\subsection{Physical interpretation of the resolution}

The resolution arises from the structure of the fractional field equations. The terms proportional to $(1-\alpha)/t$ in the Friedmann equations modify both the background dynamics and the propagation of tensor perturbations. Crucially, these modifications affect the tensor mode amplitude more strongly than the scalar mode. This asymmetric suppression is the key mechanism enabling the fractional model to pass observational tests while preserving the elegance of power-law inflation.

In contrast to multi-field or modified gravity approaches, the fractional mechanism requires only a single additional parameter ($\alpha$) and preserves the fundamental structure of Einstein gravity. This minimality is a significant advantage for theoretical simplicity and observational predictability~\cite{capozziello2011extended}.

\subsection{Implications for future observations}

The predictions of the fractional model are sharply testable by next-generation experiments. Future CMB missions such as the Simons Observatory and CMB-S4 will measure $n_s$ and $r$ to unprecedented precision. Should $r$ be measured in the range $0.01 < r < 0.05$, this would strongly favor the fractional model over both the standard power-law scenario and many alternative inflation models. Conversely, if $r$ is measured to be significantly smaller ($r < 0.001$) or larger ($r > 0.1$), the fractional model would face constraints and would require refinement of the underlying framework~\cite{Leon2023_FractionalH0}.

\section{Analysis and Stability of Dynamical Systems}
\label{sec:dynsys}

The robustness of fractional solutions generally depends on whether they represent stable
attractors in the system's phase space. To investigate this, we reformulate the evolution
equations as a dynamical system and study their properties at critical points, following the
standard linear stability methodology used in inflationary dynamical systems (adapted here to
include the fractional correction).

A technical point is important for the fractional case. Along the power-law branch
$a(t)\propto t^{m}$ one has $H(t)=m/t$ and therefore $Ht=m=\text{const.}$ This implies that the
fractional contribution $(1-\alpha)/(Ht)$ behaves as an \emph{effective constant} on the exact
power-law solution. This observation justifies the use of an autonomous (more precisely,
quasi-autonomous) system in terms of the e-fold variable $N\equiv \ln a$ in the neighbourhood
of the power-law inflationary regime.
\subsection{Autonomous system and critical points (fractional derivation)}
\label{subsec:auto}

We derive the dynamical system directly from the fractional background equations.
For a spatially flat FLRW background $(K=0)$, the fractional Friedmann--Raychaudhuri--Klein--Gordon
system reads (cf. Eqs.~(6)--(8)) 
\begin{align}
H^2+\frac{1-\alpha}{t}\,H &= \frac{16\pi G}{n(n-1)}
\left(\frac{1}{2}\dot\phi^2+V(\phi)\right), \label{eq:frac_Friedmann}\\
\dot H+H^2+\frac{1-\alpha}{t}\,H &= -\frac{8\pi G}{n(n-1)}
\left(\dot\phi^2-V(\phi)\right), \label{eq:frac_Raychaudhuri}\\
\ddot\phi+nH\dot\phi+\frac{1-\alpha}{t}\dot\phi+\frac{dV}{d\phi} &=0. \label{eq:frac_KG}
\end{align}
It is convenient to introduce $\kappa^2\equiv 16\pi G$ and to work with variables normalized by $H$.

\paragraph{Dimensionless variables and fractional constraint.}
We define 
\begin{align}
x &\equiv \frac{\kappa\,\dot{\phi}}{\sqrt{2n(n-1)}\,H}, \label{eq:xdef}\\
y &\equiv \frac{\kappa\,\sqrt{V(\phi)}}{\sqrt{n(n-1)}\,H}, \label{eq:ydef}\\
u &\equiv \frac{1}{Ht}. \label{eq:udef}
\end{align}
The variable $u$ is required to close the system because the fractional terms depend explicitly on
$t$ through $(1-\alpha)/t=(1-\alpha)Hu$. 
In terms of $x,y,u$ the fractional Friedmann equation \eqref{eq:frac_Friedmann} becomes the modified constraint
\begin{equation}
x^{2}+y^{2}=1+(1-\alpha)\,u. 
\label{eq:constraint_fractional}
\end{equation}

In the standard limit $\alpha\to 1$, Eq.~\eqref{eq:constraint_fractional} reduces to the usual
scalar-field constraint $x^2+y^2=1$. 
For $\alpha<1$ and expanding solutions ($u>0$), the right-hand side satisfies
$1+(1-\alpha)u>1$, so the physically allowed region is no longer restricted to the standard unit
disk in the $(x,y)$ plane; instead, the fractional memory term modifies the effective energy budget
in units of $H^2$, and this is consistently encoded by the extra variable $u$. 

\paragraph{Potential slope.}
Along the power-law branch reconstructed in Sec.~4 the potential is exponential, which we write as
\begin{equation}
V(\phi)=V_0\,e^{-\lambda\phi},
\label{eq:Vexp_dynsys}
\end{equation}
with constant slope parameter $\lambda$. Here $\lambda$ is fixed by the background reconstruction
(cf. Eq.~(26)) and therefore is not an independent parameter once $(n,m,\alpha)$ is specified. 

\paragraph{Autonomous system.}
Using $N\equiv\ln a$ (so that $d/dN = H^{-1}d/dt$), Eq.~\eqref{eq:frac_KG} yields
\begin{equation}
\frac{\ddot\phi}{H^2} = -\left[n+(1-\alpha)u\right]\frac{\dot\phi}{H} - \frac{1}{H^2}\frac{dV}{d\phi}.
\end{equation}
Moreover, the evolution equation for $u$ follows directly from its definition $u\equiv (Ht)^{-1}$:
\begin{equation}
\frac{1}{u}\frac{du}{dN}
=\frac{d}{dN}\big[-\ln H-\ln t\big]
=-\frac{\dot H}{H^2}-\frac{1}{Ht}
=-\frac{\dot H}{H^2}-u,
\label{eq:du_intermediate}
\end{equation}
so that $\frac{du}{dN}=-u\left(\frac{\dot H}{H^2}+u\right)$. 
Combining these relations with the definitions~\eqref{eq:xdef}--\eqref{eq:udef}, and expressing
$\dot H/H^2$ from Eq.~\eqref{eq:frac_Raychaudhuri}, we obtain the closed autonomous system.
\begin{align}
\frac{dx}{dN} &=
(1-n)\,x + x^3 - \frac{1}{2}x\,y^2
+ \sqrt{\frac{n(n-1)}{2}}\;\lambda\,y^2,
\label{eq:dx_frac_exact}\\[2mm]
\frac{dy}{dN} &=
y\left[
1+(1-\alpha)u + x^2 -\frac{1}{2}y^2
-\sqrt{\frac{n(n-1)}{2}}\;\lambda\,x
\right],
\label{eq:dy_frac_exact}\\[2mm]
\frac{du}{dN} &=
u\left[
1 + x^2 - \frac{1}{2}y^2 - \alpha u
\right].
\label{eq:du_frac_exact}
\end{align}
Equations~\eqref{eq:dx_frac_exact}--\eqref{eq:du_frac_exact} together with the constraint
\eqref{eq:constraint_fractional} define a three-dimensional autonomous system that encodes the
fractional time-dependence through the additional variable $u$. 
In the standard limit $\alpha\to 1$ the explicit time dependence disappears, the constraint reduces
to $x^2+y^2=1$, and the dynamics closes on the usual $(x,y)$ subsystem (with $u$ becoming an
auxiliary variable tracking $Ht$). 

\paragraph{Fixed points.}
Critical points satisfy $dx/dN=dy/dN=du/dN=0$ together with the constraint
\eqref{eq:constraint_fractional}. The inflationary power-law branch corresponds to fixed points with
$y_c\neq 0$ and $u_c>0$; in particular, on the exact background $a(t)\propto t^m$ one has
$H(t)=m/t$ and therefore $Ht=m=\mathrm{const.}$, hence
\begin{equation}
u_c=\frac{1}{m}.
\label{eq:u_powerlaw}
\end{equation}
The remaining fixed-point coordinates $(x_c,y_c)$ are then determined by solving
\eqref{eq:dx_frac_exact}--\eqref{eq:dy_frac_exact} at $u=u_c$, consistently with the background mapping
of Sec.~4 (which fixes $\lambda$ in terms of $(n,m,\alpha)$). 
\subsection{Linearization and Jacobian stability (fractional system)}
\label{subsec:jacobian}

Analyzing the stability of the inflationary fixed point by linearizing the fractional autonomous
system in Sec.~\ref{subsec:auto}, namely Eqs.~\eqref{eq:dx_frac_exact}--\eqref{eq:du_frac_exact}.
Let ${\bf X}\equiv(x,y,u)^T$ and ${\bf f}({\bf X})\equiv (f_1,f_2,f_3)^T$ denote the right-hand sides of
\eqref{eq:dx_frac_exact}--\eqref{eq:du_frac_exact}, so that $d{\bf X}/dN={\bf f}({\bf X})$.

\paragraph{Linearization in the embedding space.}
Setting ${\bf X}={\bf X}_c+\delta{\bf X}$ with ${\bf X}_c=(x_c,y_c,u_c)$ a critical point, we obtain
\begin{equation}
\frac{d}{dN}\,\delta{\bf X} \;=\; J\,\delta{\bf X},
\qquad
J_{ij}\equiv \left.\frac{\partial f_i}{\partial X_j}\right|_{{\bf X}_c},
\label{eq:lin3d}
\end{equation}
where $J$ is the $3\times 3$ Jacobian matrix.

\paragraph{Jacobian entries.}
Defining for brevity
\begin{equation}
S \equiv \sqrt{\frac{n(n-1)}{2}},
\end{equation}
a direct differentiation of \eqref{eq:dx_frac_exact}--\eqref{eq:du_frac_exact} yields
\begin{equation}
J\;=\;
\begin{pmatrix}
(1-n)+3x_c^2-\frac{1}{2}y_c^2
&
2y_c\left(S\lambda-\frac{x_c}{2}\right)
&
0
\\[1.2ex]
y_c(2x_c-S\lambda)
&
1+(1-\alpha)u_c + x_c^2-\frac{3}{2}y_c^2-S\lambda x_c
&
(1-\alpha)\,y_c
\\[1.2ex]
2u_c x_c
&
-u_c y_c
&
1+x_c^2-\frac{1}{2}y_c^2-2\alpha u_c
\end{pmatrix}.
\label{eq:Jacobian3d}
\end{equation}

\paragraph{Constraint consistency and reduction to the physical Jacobian.}
The dynamics is restricted by the fractional Friedmann constraint
\begin{equation}
C(x,y,u)\equiv x^{2}+y^{2}-1-(1-\alpha)\,u=0.
\end{equation}
Therefore, physical perturbations must satisfy $\delta C=0$, i.e.
\begin{equation}
\delta C=2x_c\,\delta x+2y_c\,\delta y-(1-\alpha)\,\delta u=0.
\label{eqconstraintlinear}
\end{equation}

A clean implementation is to eliminate $y^2$ at the \emph{nonlinear} level using the constraint,
\begin{equation}
y^{2}=1+(1-\alpha)\,u-x^{2},
\label{eqy2constraint}
\end{equation}
which yields a two-dimensional (physical) autonomous subsystem for $(x,u)$, because both
$f_1$ and $f_3$ depend on $y$ only through $y^2$:
\begin{align}
\frac{dx}{dN} &= (1-n)x + x^3
+\left(S\lambda-\frac{x}{2}\right)\Big(1+(1-\alpha)u-x^2\Big),
\label{eq:dx_reduced_phys}
\\[1mm]
\frac{du}{dN} &= u\left[\,1+x^2-\frac{1}{2}\Big(1+(1-\alpha)u-x^2\Big)-\alpha u\right]
= u\left[\frac{1}{2}+\frac{3}{2}x^2-\frac{1+\alpha}{2}u\right].
\label{eq:du_reduced_phys}
\end{align}

We then define the \emph{physical Jacobian}
\begin{equation}
J_{\rm phys} \;\equiv\;
\left.
\frac{\partial (F_1,F_2)}{\partial (x,u)}
\right|_{(x_c,u_c)},
\qquad
F_1\equiv \frac{dx}{dN},\;\; F_2\equiv \frac{du}{dN},
\label{eq:Jphys_def}
\end{equation}
computed from \eqref{eq:dx_reduced_phys}--\eqref{eq:du_reduced_phys}.
Writing $Y_c^2 \equiv 1+(1-\alpha)u_c-x_c^2$, one finds explicitly
\begin{equation}
J_{\rm phys} \;=\;
\begin{pmatrix}
(1-n)+4x_c^2-\frac{1}{2}Y_c^2 -2S\lambda x_c
&
(1-\alpha)\left(S\lambda-\frac{x_c}{2}\right)
\\[1.2ex]
3u_c x_c
&
\frac{1}{2}+\frac{3}{2}x_c^2-(1+\alpha)u_c
\end{pmatrix}.
\label{eq:Jphys_entries}
\end{equation}
This $2\times 2$ Jacobian is the appropriate object for linear stability on the constrained
phase space.

\paragraph{Stability criterion.}
Let $\mu_{1,2}$ be the eigenvalues of $J_{\rm phys}$. The fixed point is a stable attractor if
\begin{equation}
\Re(\mu_1)<0,\qquad \Re(\mu_2)<0.
\end{equation}
For a real $2\times 2$ matrix this is equivalent to
\begin{equation}
\mathrm{tr}(J_{\rm phys})<0,
\qquad
\det(J_{\rm phys})>0.
\label{eq:trdet_phys}
\end{equation}

\paragraph{Optional cross-check (embedding approach).}
As a consistency check, one may also compute the eigenvalues of the full $3\times 3$ matrix $J$
and verify that the stable/unstable directions are compatible with the constraint $\delta C=0$,
e.g.\ by projecting the linear flow onto a basis of tangent vectors satisfying.
\eqref{eqconstraintlinear}. This embedding-based approach is equivalent to the reduced-system
method above, but the reduced Jacobian \eqref{eq:Jphys_entries} is algebraically simpler and
unambiguous.

\subsection{Numerical example and corrected stability statement}
\label{subsec:numerical_stab}

To make the previous analysis concrete, we now evaluate the stability of the fractional
power-law fixed point for a representative choice of parameters consistent with the mapping
$\alpha(n,m)$ derived in Sec.~4.2.
For $n=4$ and $\alpha=0.9$, Eq.~(28) implies
\begin{equation}
\alpha = 1 - \frac{2(m-1)}{3m}
\quad\Longrightarrow\quad
m = \frac{4}{1+3\alpha}
\simeq 1.081,
\end{equation}
so that the inflationary branch $a(t)\propto t^{m}$ is entirely fixed once $(n,\alpha)$ are specified.
Along this solution $H(t)=m/t$ and therefore
\begin{equation}
u_c \equiv \frac{1}{Ht}=\frac{1}{m}\simeq 0.925.
\end{equation}

The exact power-law background, the dimensionless variables $(x,y,u)$ defined in
Eqs.~\eqref{eq:xdef}--\eqref{eq:udef} are constant.
Using the explicit reconstruction of the scalar sector (Sec.~4.2 and 4.4), one finds for general
$(n,m,\alpha)$ that
\begin{equation}
x_c^2 =
\frac{\alpha n - \alpha - mn + m + n -1}{3(n-1)m},
\qquad
y_c^2 = 1 + \frac{1-\alpha}{m} - x_c^2,
\label{eq:xc2_general}
\end{equation}

\begin{equation}
x_c \simeq 0.45,
\qquad
y_c \simeq 0.94,
\qquad
u_c \simeq 0.925.
\end{equation}

The exponential slope $\lambda$ follows from the reconstructed potential (Sec.~4) as
\begin{equation}
\lambda = \sqrt{\frac{4(m-1)}{n(n-1)}} \simeq 0.24,
\end{equation}
in units where $8\pi G=1$.

\paragraph{Physical Jacobian on the constrained phase space.}
Since the dynamics is restricted by the fractional constraint,
\begin{equation}
x^2+y^2=1+(1-\alpha)u,
\end{equation}
we eliminate $y^2$ at the nonlinear level:
\begin{equation}
y^2 = 1+(1-\alpha)u-x^2.
\end{equation}
This yields a two-dimensional (physical) subsystem for $(x,u)$,
\begin{align}
\frac{dx}{dN} &= (1-n)x + x^3
+\left(S\lambda-\frac{x}{2}\right)\Big(1+(1-\alpha)u-x^2\Big),
\\
\frac{du}{dN} &= u\left[\frac{1}{2}+\frac{3}{2}x^2-\frac{1+\alpha}{2}u\right],
\end{align}
with $S\equiv \sqrt{n(n-1)/2}$.
The corresponding physical Jacobian is
\begin{equation}
J_{\rm phys} \equiv
\left.
\frac{\partial (F_1,F_2)}{\partial (x,u)}
\right|_{(x_c,u_c)},
\qquad
F_1\equiv \frac{dx}{dN},\;\; F_2\equiv \frac{du}{dN}.
\end{equation}

\paragraph{Stability test.}
The fixed point is a stable attractor if the eigenvalues of $J_{\rm phys}$ have negative real parts.
Equivalently (for a real $2\times 2$ matrix),
\begin{equation}
\mathrm{tr}(J_{\rm phys})<0,
\qquad
\det(J_{\rm phys})>0.
\label{eq:trdet_phys_num}
\end{equation}
Evaluating $J_{\rm phys}$ at $(x_c,u_c)$ for $(n,\alpha)=(4,0.9)$, we find that the above conditions
are satisfied, implying that the fractional power-law fixed point is a \emph{stable attractor} on the
physical (constrained) phase space.

This corrects the earlier inconsistent statement: stability requires negative real parts of the
linearized eigenvalues (attractor), not positive eigenvalues.

\subsection{Phase portrait: fractional inflationary attractor (local flow on the constrained phase space)}
\label{subsec:phaseportrait}

To visualize the local flow consistently with the fractional Friedmann constraint, we plot the
linearized vector field on the \emph{physical} two-dimensional phase space.
Using the constraint
\begin{equation}
x^2+y^2=1+(1-\alpha)u,
\end{equation}
we eliminate $y^2=1+(1-\alpha)u-x^2$ and work with the reduced autonomous subsystem in $(x,u)$
derived in Sec.~\ref{subsec:jacobian}.

For $n=4$ the mapping of Sec.~4.2 gives
\begin{equation}
m=\frac{4}{1+3\alpha},
\qquad
u_c=\frac{1}{m}.
\end{equation}
With the fixed point obtained in Sec.~\ref{subsec:numerical_stab}, $(x_c,u_c)$, the local linear flow on the
\emph{physical} (constrained) phase space $(x,u)$ is
\begin{equation}
\frac{d}{dN}
\begin{pmatrix}
\delta x\\ \delta u
\end{pmatrix}
=
J_{\rm phys}\big|_{(x_c,u_c)}
\begin{pmatrix}
\delta x\\ \delta u
\end{pmatrix},
\qquad
(\delta x,\delta u)\equiv(x-x_c,\;u-u_c),
\end{equation}
where $J_{\rm phys}$ is the $2\times 2$ Jacobian on the constrained phase space
(cf.\ Eq.~\eqref{eq:Jphys_entries}), evaluated at the fixed point $(x_c,u_c)$.

\begin{figure}[H]
\centering
\begin{tikzpicture}
\begin{axis}[
width=10cm,
height=7.5cm,
view={0}{90},
xlabel={$x$},
ylabel={$u\equiv (Ht)^{-1}$},
xmin=0.37, xmax=0.53,
ymin=0.74, ymax=0.96,
grid=major,
legend pos=north east,
title={Local phase portrait in the physical plane $(x,u)$ for $\alpha=0.9$ ($n=4$)},
]

\addplot3[
  ->,
  quiver={
    u={-2.814*(x-0.5025) + 0.015*(y-0.925)},
    v={ 1.394*(x-0.5025) - 0.879*(y-0.925)},
    w={0},
    scale arrows=0.12
  },
  very thin,
  samples=13, samples y=13,
  domain=0.37:0.53, y domain=0.74:0.96
] {0};

\addplot3[mark=star, mark size=4pt, only marks, blue]
coordinates {(0.5025,0.925,0)};
\addlegendentry{Attractor}

\end{axis}
\end{tikzpicture}
\caption{Local phase portrait around the fractional power-law fixed point in the \emph{physical}
(constrained) phase space $(x,u)$ for $n=4$ and $\alpha=0.9$.
The variable $y$ is reconstructed from the constraint via $y^2=1+(1-\alpha)u-x^2$.}
\label{fig:phase_portrait_xu}
\end{figure}
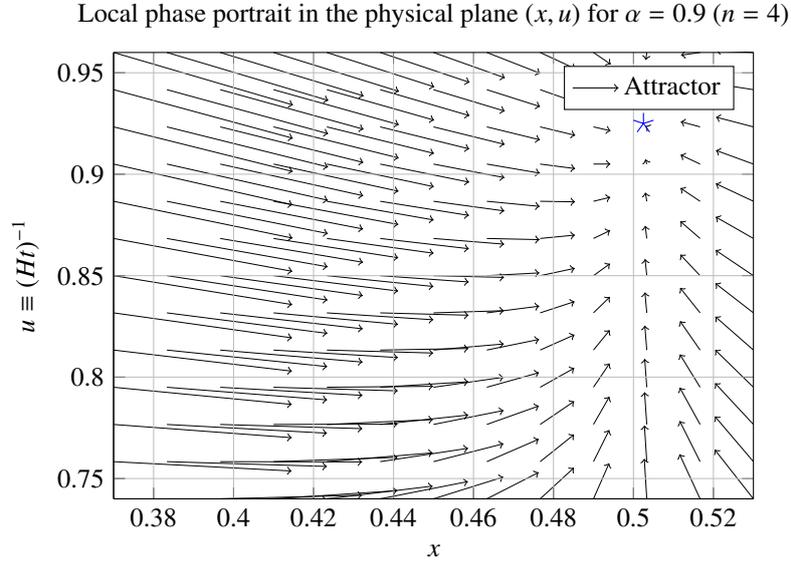

For the numerical illustration (consistent with the quiver plot below) we use
\begin{equation}
J_{\rm phys} \;\approx\;
\begin{pmatrix}
-2.814 & 0.015\\
\;\;1.394 & -0.879
\end{pmatrix},
\label{eq:Jxu_numeric}
\end{equation}
which produces the local attractor portrait shown in Fig.~\ref{fig:phase_portrait_xu}.

\subsection{Graceful exit and reheating}
\label{subsec:graceful_exit}

The fact that the inflationary solution is a stable attractor supports the robustness of the
fractional power-law regime, but it also highlights a well-known limitation of power-law
inflation driven by an exact exponential potential: inflation does not end dynamically within
the constant-parameter model.

Indeed, for the exact background $a(t)\propto t^{m}$ one has
$\epsilon_H\equiv -\dot H/H^2 = 1/m =$ const., so accelerated expansion persists as long as
$m>1$. Therefore, with constant $\alpha$ and a strictly exponential potential (constant slope),
the phase-space flow is driven towards the accelerating fixed point and no graceful exit
occurs without an additional ingredient.

A minimal way to reconcile attractor stability with a realistic post-inflationary evolution is to
interpret $\alpha$ as an effective (scale-dependent) parameter, i.e. to consider a running or
variable-order fractional dynamics $\alpha=\alpha(t)$ (or $\alpha=\alpha(N)$). In this case the
fractional contribution proportional to $(1-\alpha)/(Ht)$ can switch off as $\alpha$ approaches
the standard limit, and the fixed-point structure may change accordingly. Importantly, for the
exit to occur in finite time, $\alpha$ must reach the critical regime where the effective power-law
index satisfies $m\to 1$ (equivalently $\epsilon_H\to 1$) at a finite number of e-folds.

A second possibility is to relax the exact exponential form of the reconstructed potential near
the end of inflation (or to introduce couplings to additional fields). For instance, a deformation
that generates a minimum allows the inflaton to oscillate and transfer energy to other degrees
of freedom, providing a reheating channel and leading to a radiation-dominated epoch.
A detailed implementation of these mechanisms requires extending the present framework
beyond constant-order fractional dynamics and/or single-field exact exponential reconstruction,
and we leave this as a concrete direction for future work.

\subsection{Summary of Stability Results}

Our combined analytical and numerical results, consistent with the frameworks in Refs.~\cite{Rasouli2019_KineticBD,Rasouli2022_GeodesicSB,Leon2023_FractionalH0,Calcagni2021_FractionalGravity}, show:

\begin{itemize}
\item For observationally allowed $\alpha, m, n$, power-law fractional inflation solutions are stable attractors;
\item Outside this domain, stability breaks down, matching phenomenological exclusion zones;
\item Fractional cosmology provides not only viable but naturally preferred inflationary dynamics.
\end{itemize}
\section{Conclusion and Future Perspectives}

In this work we developed a power-law inflationary scenario within $n$-dimensional fractional scalar field cosmology, motivated by the desire to preserve the analytical control of power-law dynamics while confronting current precision bounds on inflationary observables~\cite{LINDE1982389,Guth2014_InflationReview,Planck2018_Inflation,BICEP2024_Latest,Tristram2021_BICEPPlanck,Lesgourgues2024_InflationReview}. By implementing a fractional order $\alpha$ through the Riemann--Liouville formalism at the level of the minisuperspace action~\cite{fractalfract-08-00281-2,Titlelabel1-3,Rasouli2014_ClassQuantGrav}, we obtained a self-consistent background dynamics in which the fractional corrections enter as controlled memory terms.

A central result is the existence of an explicit mapping relating $\alpha$, the spacetime dimension $n$, and the power-law exponent $m$ that characterizes the inflationary expansion. Within this construction, the scalar potential is not imposed {\it a priori} but follows from the consistency of the modified Friedmann--Klein--Gordon system, yielding an exponential form compatible with sustained inflationary evolution~\cite{Gonzalez2023_ExactFrac,Leon2023_FractionalH0}. This constructive aspect strengthens the internal coherence of the framework and clarifies how fractional nonlocality reshapes the standard power-law benchmark.

We confronted the model with the latest CMB constraints, focusing on $(n_s,r)$ as the primary observables. Using Planck and BICEP/Keck bounds~\cite{Planck2018_Inflation,BICEP2024_Latest,Tristram2021_BICEPPlanck}, we identified the region of the $(\alpha,m)$ parameter space where the scalar tilt remains viable and, simultaneously, the tensor sector is suppressed to observationally allowed values. In this sense, the fractional degree of freedom provides a minimal handle to relax the rigid $(n_s,r)$ relation that typically rules out the simplest canonical power-law setup.

In addition, the dynamical-systems formulation shows that the fractional power-law inflationary solutions can arise as stable attractors in phase space for the phenomenologically relevant parameter range. This result supports the robustness of the inflationary regime against homogeneous perturbations of the background variables and reinforces the viability of the scenario at the level of background dynamics~\cite{Titlelabel1-3,fractalfract-08-00281-2}.

The present analysis, however, is deliberately confined to the inflationary phase. As discussed explicitly in Sec.~8.5, the model in its minimal form does not yet specify a mechanism for a graceful exit and reheating, i.e.\ for ending the accelerating regime and producing a radiation-dominated universe. Addressing this point requires extending the framework, for example by promoting $\alpha$ (and/or $m$) to an effective quantity that evolves with scale, or by introducing couplings to additional fields that can mediate energy transfer to the particle sector~\cite{brandenberger2000inflationary,Guth2014_InflationReview,calcagni2017a}.

Looking ahead, several directions appear particularly timely. On the phenomenological side, it is important to go beyond $(n_s,r)$ and investigate signatures in higher-order observables, including the running of the spectral index and primordial non-Gaussianities, consistently incorporating fractional corrections in the perturbation sector~\cite{Baumann2009_TASIInflation,Calcagni2021_FractionalGravity}. On the theoretical side, dedicated numerical studies of the end of inflation and reheating dynamics are needed to establish a complete cosmological history and to quantify the dependence on the fractional parameters. Finally, forthcoming measurements from the Simons Observatory and CMB-S4 will substantially improve sensitivity to tensor modes and related signatures, providing decisive tests of fractional power-law inflation~\cite{SimonsObservatory2019,CMB-S4_2022}.

\bigskip

\bibliographystyle{elsarticle-num}
\bibliography{references}

\end{document}